# *Subnetwork hierarchies of biochemical pathways*

*Petter Holme*[1], *Mikael Huss*[2] and *Hawoong Jeong*[3]

[1]*Department of Theoretical Physics, Umeå University, 901 87 Umeå, Sweden,*
[2]*SANS, NADA, Royal Institute of Technology, 100 44 Stockholm, Sweden,*
[3]*Department of Physics, Korea Advanced Institute of Science and Technology, Taejon, 305-701, Korea.*

Dated: May 21, 2002

**ABSTRACT**

We present a method to decompose biochemical networks into subnetworks based on the global geometry of the network. This method enables us to analyse the full hierarchical organisation of biochemical networks and is applied to 43 organisms from the WIT database. Two types of biochemical networks are considered: metabolic networks and whole-cellular networks (also including e.g. information processes). Conceptual and quantitative ways of describing the hierarchical ordering are discussed. The general picture of the metabolic networks arising from our study is that of a few core-clusters centred around the most highly connected substances enclosed by other substances in outer shells, and a few other well-defined subnetworks.

## 1 INTRODUCTION

In the last few years, several studies have addressed graph-theoretical aspects of biochemical networks (see e.g. Schuster and Hilgetag, 1994; Schuster *et al.*, 1999; Jeong *et al.*, 2000; Fell and Wagner, 2000; Jeong *et al.*, 2001; Schuster *et al.*, 2002). These studies have found that many characteristics of these networks are fairly universal among both prokaryotes and eukaryotes. For example, the distribution of degree (sometimes called connectivity—in general the number of neighbours to a node in a network) has been shown to have a power-law distribution in a great variety of organisms, meaning that a few substances (ATP, $H_2O$, etc.) dominate the network structure (Jeong *et al.*, 2001). Graph theoretical methods have also been used to identify structural subnetworks (Schuster *et al.*, 2002), and even if such methods do not incorporate detailed reaction kinetic information, the obtained subnetworks are arguably biologically significant. The desire for finding subnetworks arises from the vastness of the biochemical networks; even a prokaryote such as *E. coli* has a metabolism involving over 850 substances and 1500 reactions. Attempts to elucidate e.g. a bacterium's metabolic repertoire thus face the problem of combinatorial explosion. A fundamental question is therefore what the hierarchical organisation of subnetworks looks like. Can relevant subnetworks be found at arbitrary sizes? Is it at all relevant to talk of subnetworks, or must the whole network always be taken into account? This paper aims to answer these questions by proposing a general method for partitioning a biochemical network into subnetworks by successively removing reactions of high betweenness centrality—reactions situated between areas of many interior pathways (i.e. well-defined subnetworks). Besides finding explicit subnetworks of arbitrary sizes, this method also enables us to investigate the full hierarchical organisation of a cellular network—how a subnetwork can be divided into sub-subnetworks and so on.

## 2 TRACING THE HIERARCHICAL SUBNETWORK STRUCTURE

### 2.1 Networks

We represent the metabolic network as a directed bipartite graph $G = (S, R, L)$ where $S$ is the set of nodes representing substrates, $R$ is the set of nodes representing reactions (or, following the terminology in Jeong *et al.* (2001), temporary educt-educt complexes), and $L$ is the set of directed links—ordered pairs of one node in $S$ and one node in $R$. $s_1, \cdots, s_n \in S$ is involved in a reaction $r \in R$ with products $s'_1, \cdots, s'_{n'} \in S$, if and only if $(s_1, r), \cdots, (s_n, r) \in L$ and $(r, s'_1), \cdots, (r, s'_{n'}) \in L$.

The networks we use (the same data set as in Jeong *et al.*, 2001) were constructed from the WIT database[1] (Overbeek *et al.*, 2000) consisting of 43 organisms from all domains of life—6 archae, 32 bacteria, and 5 eukaryotes. Not all of these networks are complete, and a few of the smallest subnetworks we detect are results of this incompleteness. Nevertheless, the global organisation and the larger detected subnetworks should be fairly insensitive to discrepancies in the database (Jeong *et al.*, 2000).

---
[1]Similar information can be obtained from the KEGG (Kanehisa and Goto, 2000), EcoCyc (Karp *et al.*, 2000) and EMP (Selkov *et al.*, 1996) databases.

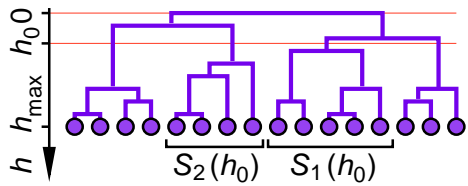

**Fig. 1:** A simple hierarchical clustering tree. A horizontal cut gives the tighter connected subgraphs below, and looser connections above. $S_i(h_0)$ is the size of the $i$'th largest connected subgraph at height $h_0$. Note that the root is at the top and $h$ grows downwards.

We distinguish between metabolic networks and whole-cellular networks—networks representing the full set of cellular pathways. WIT divides the latter into the following subcategories: intermediate metabolism and bioenergetics, information pathway, electron transport, transmembrane transport, signal transduction and structure and function of cell, of which intermediate metabolism and bioenergetics constitute the metabolic networks.

### 2.2 Decomposition algorithms

The standard set of metabolic pathways described in biological literature is sometimes too rigid to capture the essence of what is happening in an organism; the subsystems identified in this way will often to a large extent overlap and intertwine. As a complement to the traditional approach there is a need for unbiased analytic methods, such as the one proposed by Schuster *et al.* (2002), a method for decomposing a biochemical network into subnetworks based on the degree of the metabolites.[2] The idea of Schuster *et al.* is to label metabolites with degree $k$ larger than some threshold value $k_{\max}$ as 'external'—as either a source or sink, and then consider connected components of 'internal' metabolites as subnetworks, the motivation being that the system can be regarded as buffered with respect to the substrates participating in the largest number of reactions (Fell and Wagner, 2000). If the relabelling of an internal node as external is interpreted as deleting the node from the network of internal nodes, this method is equivalent to the attack vulnerability study of Albert *et al.* (2000), where scale-free networks' (such as biochemical networks') response to the removal of nodes in order of degree was discussed.

---

[2]Another potentially interesting cluster identification algorithm applied to backbone clusters of residues in proteins is presented in Patra and Vishveshwara, 2000.

The heuristic motivations and *a posteriori* success (for subnetwork detection) of the method of Schuster *et al.* makes it an important contribution to biochemical pathway analysis. A potential drawback of this method is that networks might have inherent non-local features not possible to be detected by local quantities such as degree. (Non-local effects are known to be an important factor in e.g. social networks; see Granovetter, 1973.) To be specific, consider a node $m$ with degree $k_m > k_{\max}$ neighbours, all except one having $k = 1$. This is a local centre but globally (for a large enough network) a peripheral node. Then all these $k = 1$-nodes would be classified as belonging to individual one-node clusters, while a more informative categorisation would be to have the $k = 1$ and $k = k_m$ nodes in the same cluster. Indeed, this kind of configuration actually exists, which will be shown later. The method for identifying subnetworks (presented in the following two sections) is similar to that of Schuster *et al.*, only that our algorithm removes reaction nodes rather than substrates, and the removal is based on a global centrality measure (betweenness) rather than a local (degree).

### 2.3 Constructing hierarchies of subnetworks

Structural studies of networks have long history in sociology, and many methods and concepts can be brought over to biological network studies as well. The traditional way of detecting hierarchies of subnetworks (or the "community structure") in social networks has been by hierarchical clustering methods (Johnson, 1976): First, one calculates some measure of strength of links (ties) between the nodes (persons) of the network. (In biochemical networks a good measure of strength would be the mass flux between two substances.) Then one reconstruct the network by adding the links in order of this strength. In this way one can construct a hierarchy tree, where the tightest connected subgraphs are joined by links close to the root of the tree, and the most long-range inter-community links are close to the root. (See Fig. 1.)

However, these hierarchical clustering methods have some inherent flaws. For example, just as by Schuster's method above, nodes with one neighbour often become classified as belonging to a one-node cluster. In remedy, Girvan and Newman (2002) proposed an elegant method where one deconstructs the network by successively deleting links of highest betweenness centrality (see next section)—roughly speaking links carrying many shortest paths—which are likely to lie between tightly connected subnetworks.

### 2.4 Betweenness centrality

For an undirected graph the betweenness centrality $C_B$ (Freeman, 1977)—or for short, betweenness—of a node $v$ is the number of shortest paths between pairs of nodes that pass $v$ (if more than one shortest path exist between $u$ and $u'$ passes $v$, the fraction of shortest paths through $v$ contributes to its betweenness). In this way, the betweenness measure finds the communication centres of the networks. In a tightly connected subnetwork the shortest paths are relatively direct and do not converge at specific nodes; thus the betweenness within a tightly connected area is relatively low, whereas nodes in less connected areas between tightly connected subgraphs have a higher betweenness. For the purposes of this work we are interested in reaction nodes that are central for paths between metabolites or other molecules; thus we redefine the betweenness centrality of reaction nodes as follows—for $r \in R$:

$$C_B(r) = \sum_{m \in M} \sum_{m' \in M \setminus \{m\}} \frac{\sigma_{mm'}(r)}{\sigma_{mm'}}, \quad (1)$$

where $\sigma_{mm'}(r)$ is the number of shortest paths between $m$ and $m'$ that passes through $r$, and $\sigma_{mm'}$ is the total number of shortest paths between $m$ and $m'$. For calculating betweenness we use the fast algorithm by Brandes (2001).

A networks' response to the successive deletion of central nodes and links, as in the above algorithm, is termed attack vulnerability (Albert *et al.*, 2000) in network literature. Deletion in order of recalculated betweenness has been proven particularly effective in some cases (Holme *et al.*, 2002). It should be noted that reactions of high betweenness are central in the wiring of a biochemical network, but not necessarily in the network itself since reaction coefficients and concentrations are not taken into account (and should not be—then high centrality would not correspond to being situated between highly connected areas).

### 2.5 Our algorithm

As we represent the biochemical network as a bipartite graph (so that all substances—metabolites, macromolecules, complexes etc.—are separated by reaction nodes and an even number of links), we modify the algorithm of Girvan and Newman and successively delete reaction nodes with high betweenness with respect to substrates/products and enzymes. A reaction corresponds to a passage through all inwards links to a reaction node. To take this into account we regard the effectiveness of a link in the betweenness definition as proportional to $1/k_{in}(r)$ ($k_{in}(r)$ being the in-degree, or number of substrates to a reaction). The rescaled, effective betweenness thus becomes:

$$c_B(r) = C_B(r)/k_{in}(r) \quad (2)$$

With these definitions the algorithm comes closer to that of Girvan and Newman (2002), where link betweenness is the operative centrality measure—$c_B(r)$ could also be interpreted as the average link-betweenness of $r$'s inward links.

With the above modifications, the algorithm consists of the following steps repeated until no reaction nodes remain:

1. Calculate the effective betweenness $c_B(r)$ for all reaction nodes.

2. Remove the reaction node with highest effective betweenness and all its in- and out-going links.

3. Save information about the current state of the network (such that how many clusters there are, and what nodes that belongs to a specific cluster).

If many reaction nodes have the same betweenness in step 2, we remove all of them at once. (In this case the time scale of the algorithm is relevant one would have to find some condition to choose only one.) To speed up the algorithm, only reaction nodes belonging to the cluster of the last deletion need to be considered for recalculation of the betweenness. The information obtained in step 3 is used to construct the hierarchy trees and statistics about the ordering.

## 3 RESULTS

### 3.1 General shape of the hierarchy trees

As example of a hierarchical clustering trees, Fig 3, of metabolic and whole-cellular networks we consider those of the bacterium *T. pallidum*.[3] (*T. pallidum* is the pathological agent of syphilis. A recent review of its functions in a genomic perspective is given in Norris *et al.*, 2001.) Most constituents are connected into a giant component (a cluster whose size scales linearly with the total number of nodes, see e.g. Janson *et al.*, 2000). Close to the root (i.e. in the top of Fig. 3) the giant component is still existent, but at height $h \approx 0.8\,h_{\max}$ of the hierarchy tree the giant component starts to break into well-defined clusters. When a cluster breaks in sub-clusters of similar sizes we say the hierarchy tree has "community-type ordering" at the level in question (see Fig. 2a)). When the a cluster breaks into one large sub-cluster and many isolated nodes the level has "shell-type ordering" (see Fig. 2(b)). These concepts designed to signify the extreme cases of a most or least symmetric splitting of the clusters, intermediate cases where the cluster splits into sub-clusters of various sizes can of course also occur. In real hierarchy trees, shell-type ordering is frequent in the whole tree, and dominates the levels closest to the root (with small $h$). Community-type ordering, on the other hand, is only frequent at high $h$. Still, community-type ordering is needed to demarcate well-defined subnetworks. This lack of a community-type ordering close to the root of the tree (such as seen in e.g. social networks and ecological food-webs in Girvan and Newman, 2002) is related to the highly heterogeneous centrality distribution (Jeong *et al.*, 2001) of cellular biochemical networks: The giant component is tightly connected by the many paths involving the most connected substances, ATP, NADH, $H_2O$, and so on. The core of the metabolism is centred around these most connected substances, hence most well-defined subnetworks must contain these, but this is precisely to say that the subnetwork containing these is sure to dominate most levels of the organisation. Community-type ordering occurs when either none of these most connected substances is central in some subnetwork, or when these substances fall into different subnetworks.—Both these cases occur in e.g. *T. pallidum* (see Fig. 3; a functional description of these subnetworks will be given later): *N*-acetyl-D-glucos-amine 1-phosphate, D-glucosamine 1-phosphate, di-hydrolipoamide, *S*-acetyldihydrolipoamide, CoA, and acetyl-CoA define small subnetwork not including any of the most connected substances (see Fig 6). The highly connected orthophosphate and the substances most tightly connected to it (*α*-D-ribose 1-phosphate, *α*-D-ribose 1-phosphate adenine, adenosine, hypox-anthine, and inosine) define another subnetwork that at a higher level of organisation (lower *h*) is joined by other substances (2-deoxy-D-ribose 1-phosphate, deoxyadenosine, guanine, and guanosine) to a more loosely connected subnetwork. The general picture that arises from the study of hierarchy trees is thus that the cellular biochemical networks consist of outer shells encapsulating a core of the most connected substances, with a few well-defined subnetworks at an intermediate level of the organisation. This picture is (more or less) the same for all the 43 organisms examined.

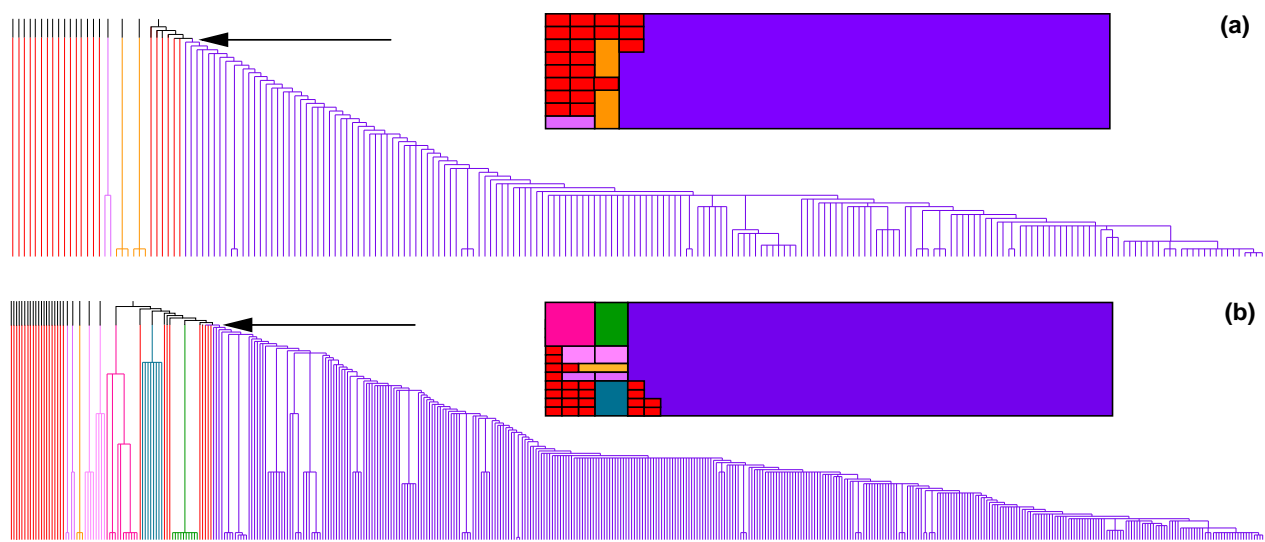

**Fig. 3:** The hierarchical clustering trees of *T. pallidum*. (a) shows the tree for the metabolic network, (b) shows the whole-cellular network. The squares represent the subnetwork configuration at $h = 0.1\,h_{\max}$ (the height indicated by the arrow). Sizes of the squares are proportional to the size of the clusters they represent.

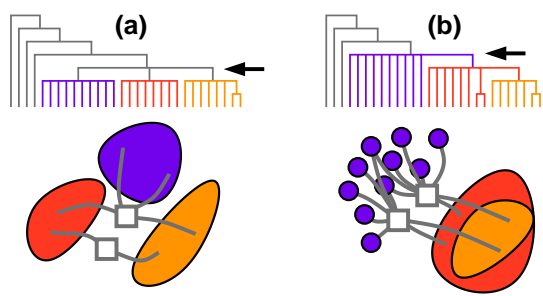

**Fig. 2:** Schematic picture of the two different orderings in hierarchy trees. (a) Community-type ordering—same level core-clusters connected by outer parts of the network. (b) Shell-type—a sequence of core-clusters contained in each other. The squares symbolises the reaction nodes that is deleted at the height marked by the arrow. In (a) three subnetworks of similar sizes gets disconnected when the reaction odes are removed. In (b) many individual metabolite nodes (circles) gets isolated.

---

[3] Hierarchical clustering trees of the other 43 organisms of the WIT database (Overbeek *et al.*, 2000) used in this study can be seen on http://www.tp.umu.se/forskning/networks/meta/.

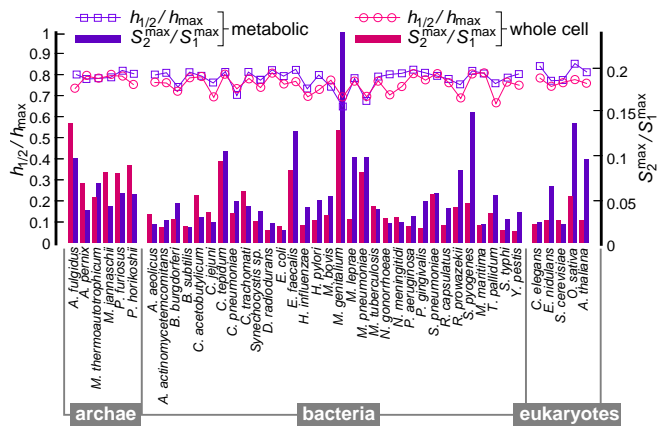

**Fig. 4:** The relative size of the network $N/N_{max}$; the ratio between the largest values of the second largest and largest connected subgraphs $S_2^{max}/S_1^{max}$; and the relative half-height $h_{1/2}/h_{max}$ for the 43 studied organisms.

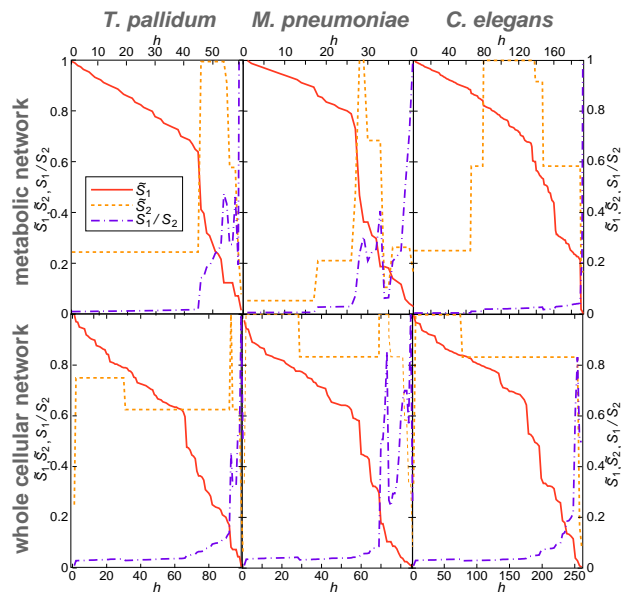

**Fig. 5:** The size of the two largest subgraphs rescaled by their largest values, $\tilde{S}_1 = S_1/S_1^{max}$ and $\tilde{S}_2 = S_2/S_2^{max}$ and the ratio $S_2/S_1$; all as functions of the height of the hierarchy tree, or event time, $h$. The data is for both metabolic and whole cellular networks of the bacteria *T. pallidum* and *M. pneumoniae*, and the nematode *C. elegans*.

### 3.2 Statistics of the hierarchical ordering

An immediate impression from looking at the hierarchy trees of the 43 WIT organisms is that they are similar in the large scale and more diverse locally. Furthermore, the shell-type ordering dominates much of the small-*h* region of metabolic network, whereas in the whole cell networks, community-type ordering also occurs close to the root. This section aims to quantify these observation. A third observation is that, as *h* increases, the tree often splits in a non-uniform way, so that only individual nodes or smaller clusters are removed from the largest connected component at a time.

The large-scale shape of the tree can be measured in many ways. One simple and informative quantity is the half-height of the largest cluster $h_{1/2}$, i.e. the height *h* where the size of the largest cluster $S_1$ has decreased to half of its original value. (For notations see Fig. 1.) If uniform ordering, where clusters break into cluster of similar sizes, would dominate all levels of organisation (such as the examples from sociology and ecology in Girvan and Newman, 2002) we expect a very small relative half-height $h_{1/2}/h_{\max}$ ($h_{1/2} \propto \log h_{\max}$). In Fig. 4 the relative height of the tree where $S_1$ has decreased to half of its original value, $h_{1/2}/h_{\max}$ is displayed. Averaged over all 43 organisms this happens at $h = (0.79 \pm 0.04) h_{\max}$ for the metabolic networks and $h = (0.76 \pm 0.04) h_{\max}$ for the whole-cellular networks—a very narrow region suggesting an universal behaviour (see Fig. 4). Even though the almost constant $h_{1/2}/h_{\max}$ is not trivially related to other universal features, such as the constant average shortest-path length (Jeong *et al.*, 2001), it is consistent with the general picture of a great diversity of organisms having a very similar large-scale organisation of the biochemical pathways. In Fig. 5 the time evolution of the size of the largest cluster $S_1$ is shown for three different organisms, all having a similar convex shape of the curve of $S_1$ as a function of the height $h$ ($h$ is also the number of iterations of the algorithm where the number of connected subgraphs has increased, thus defining an event time scale).

To measure the magnitude of community-type ordering we study the $h$-evolution of the size of the second largest cluster $S_2$. In Fig. 4 $S_2^{\max}/S_1^{\max}$ is displayed for all 43 organisms. A large value of this quantity means that the network at some time has at least two subnetworks of a large and similar size and thus a pronounced community-type order. Networks with a high $S_2^{\max}/S_1^{\max}$ in fact also have a high $S_3^{\max}/S_1^{\max}$ and so on, so this quantity works well as a measure of the degree of community-type order. Fig. 4 shows that there is a large variance in $S_2^{\max}/S_1^{\max}$ (with $S_2^{\max}/S_1^{\max} = 0.06 \pm 0.05$ for the metabolic networks and $S_2^{\max}/S_1^{\max} = 0.05 \pm 0.03$ for the whole-cellular networks). Although larger databases would be needed to obtain statistical certainty $S_2^{\max}/S_1^{\max}$ for orders the organisms as archae > bacteria > eukaryotes. It is interesting to note that eukaryotes have the lowest value. A more uniform (small $S_2$) organisation is more robust, which implies that eukaryote biochemical networks are more robust than those in bacteria and archae.

In Fig. 5 we display the event time evolution of $S_1$, $S_2$, and $S_2/S_1$. In all substances $S_1$ is slowly decreasing until the sharp drop around $h_{1/2}$. In many cases $S_2$ has a narrow peak around the drop in $S_1$ (like in Fig. 5(a) and (b)), whereas in a few other cases the $S_2$-peak is very broad and a large $S_2$ exists long before the drop in $S_1$ (see Fig. 5(c)). Above this drop of $S_1$ the ratio $S_2/S_1$ quickly approaches unity. The $S_2$-peak seems to be the height where most meaningful subnetworks can be found—closer to the root the giant component is largely intact, while for the highest $h$ the subnetworks are uninterestingly small.

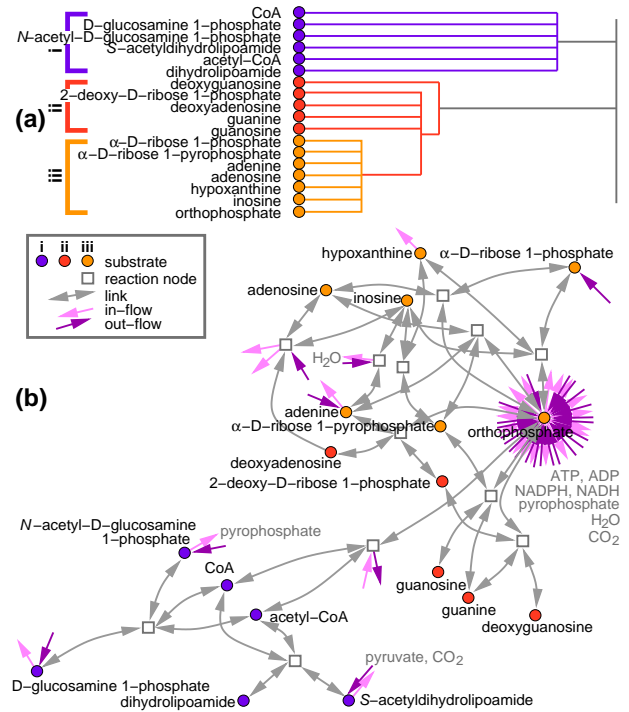

**Fig. 6:** Subnetworks from the metabolic networks of *T. pallidum*. (a) shows the part of the hierarchy tree that (b) corresponds to. Grey substance names show where shortest paths to the hubs (most connected substances of the network) enter.

### 3.3 Detected subnetworks

To give an explicit example how community- and shell-ordering are manifested in a metabolic network, we consider two small subnetworks at $h = 40$ of the hierarchy tree of *T. pallidum*'s metabolic network (Fig. 3) shown in Fig. 6. These subnetworks contains reactions associated with purine metabolism and pyruvate/acetyl-CoA conversion. The pyruvate-acetyl-CoA part ((i) in the hierarchy tree, Roman numbers refer to Fig. 6(a)) is a tightly interconnected, fairly independent subnetwork, while the purine metabolism part consists of an outer shell (ii) encapsulating a smaller core (iii), which is centred around orthophosphate and has to do with interconversions between adenosine and related nucleosides. Deoxyadenosine ends up in the outer shell (ii) while adenosine in located in the core (iii) because there are two reactions involving adenosine and orthophosphate, but only one with deoxyadenosine and orthophosphate. Overall, however, the subnetworks in this case can be readily seen to represent metabolic processes of increasing homogeneity.

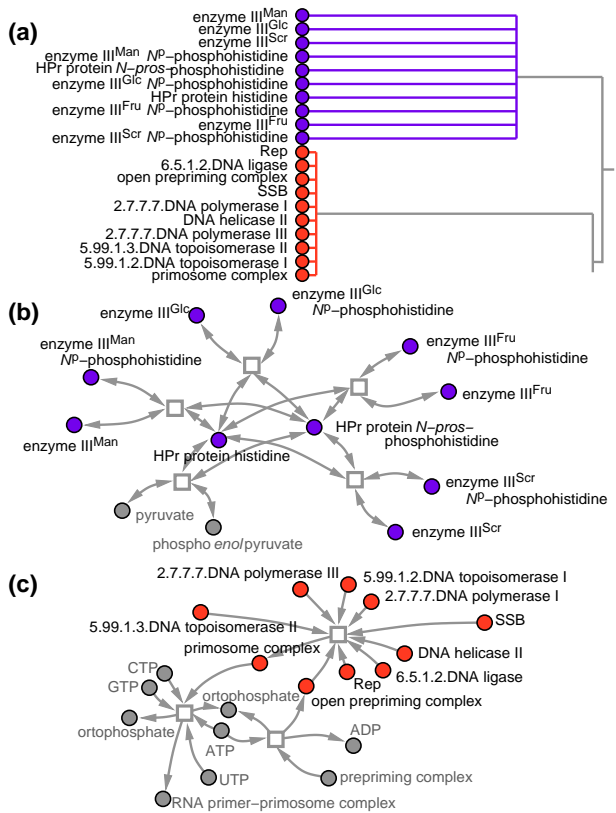

**Fig. 7:** An excerpt of the hierarchy tree (a) of *M. pneumoniae* and the corresponding subnetworks associated with sugar import (b) and DNA replication (c). Symbols are the same as in Fig. 6, except that nodes adjacent to the two clusters are marked with grey circles. Of the grey nodes' links, only those between the grey node and the cluster are shown. One cluster is three reactions away from the other by, for example, the reaction phospho*enol*pyruvate + ADP ⟶ pyruvate + ATP.

We exemplify whole-cellular networks of the bacterium *M. pneumoniae*, a bacterium causing respiratory tract infections (see the review Principi and Esposito (2001); or Ancel *et al.* for a graph-theoretical epidemics study) in Fig. 7. In general the whole-cellular networks are, perhaps not surprisingly, even more functionally distinct than the metabolic networks. One of the subnetworks (Fig. 7(b)) is a part of the bacterial phosphotransferase system (PTS), the function of which is to import carbohydrates into the cell (see Saier (2001) for an overview of the PTS). Enzyme III is an old collective name for what is now called enzyme IIA and enzyme IIB. Each of these enzymes is specific for a certain kind of carbohydrate; in Fig. 7(b), we see enzymes specific for mannitol, glucose, sucrose and fructose, respectively. The other network (Fig. 7(c)) has to do with DNA replication. (See e.g. Lewin 1999 for an overview of DNA replication.) Although both of these subnetwork differ from the metabolic networks in that the nodes are not metabolites which are interconverted but rather enzymes that interact with or are part of macromolecular complexes, they both nevertheless represent biologically meaningful groups of substances. Furthermore, the DNA replication subnetwork is centred around a reaction node with high degree (local centrality), but relatively low betweenness (global centrality). Thus local, degree-based, algorithms would have difficulties identifying such a subnetwork. Note that the subnetwork of Fig. 7(b) is ordered higher (is connected at a lower $h$) in the hierarchy than that in Fig. 7(c) since the reversibility of the reactions in Fig. 7(b) increases their betweenness.

## 4 SUMMARY AND DISCUSSION

We propose an algorithm for decomposing biochemical networks into subnetworks based on the global network structure. The algorithm—a development of the algorithms by Girvan and Newman (2002), and Schuster *et al.* (2002)—is purely graph theoretical and uses no biological criteria (cf. Schilling and Palsson, 2000). The data we study are the sets of metabolic and whole-cellular networks of 43 organisms (archae, bacteria and eukaryotes) from the WIT database. We emphasise the study of hierarchy trees to get a general view on the organisation of subnetworks. To characterise the hierarchical organisation we introduce (to the biochemical network studies) the concepts of community- and shell-ordering and quantitative measures (the relative half-height $h_{1/2}/h_{\max}$, and the relative largest size of the second largest cluster $S_2^{\max}/S_1^{\max}$).

The large-scale shape of biochemical network trees is conspicuously uniform among organisms as manifested in an universal relative half-height ($h_{1/2} = (0.79 \pm 0.04)\, h_{\max}$ for the metabolic networks and $h_{1/2} = (0.76 \pm 0.04)\, h_{\max}$ for the whole-cellular networks). The spread in $S_2^{\max}/S_1^{\max}$ shows that community-ordering is much more pronounced in some organisms than in others. A small $S_2^{\max}/S_1^{\max}$ implies a more robust network, which means that it is a quantity of potential interest for evolutionary studies (when the databases reach a size where sufficiently good statistics can be generated).

Well-defined subnetworks occur at different levels of organisation (at different heights in the hierarchy tree). This is a strong argument for looking at the whole hierarchy tree rather than the subnetwork configuration at a specific level. For the metabolic networks, the dominating structure at most levels of organisation is the largest connected component, which means that in some contexts it might be deceptive to generalise properties of subnetworks to the whole network. For the whole-cellular network, non-metabolic subnetworks, such as those representing information pathways, signal transduction and the like, are often branched of from the metabolic circuitry close to the root of the hierarchy tree. Still, the largest component is dominant over a large portion of the tree's levels. The general picture of the hierarchical organisation that emerges from our study is thus that biochemical networks have individual core-clusters of the most connected substances and their closest related substances; the rest of the substances are then organised as a more and more loosely connected outer shell, with the exception of some well-defined clusters at intermediate levels.

**ACKNOWLEDGEMENTS**

The authors thanks Prof. P. Minnhagen for fruitful discussions. P.H. acknowledges partial support from the Swedish Natural Research Council through Contract No. F5102-659/2001. H.J. acknowledges financial support from the Ministry of Information and Communication of Korea through IMT2000-B3-2. Drawing of the figures was aided by the Pajek package (http://vlado.fmf.uni-lj.si/pub/networks/pajek/).